\documentclass[format=sigconf,screen=true,nonacm=true]{acmart}

\usepackage[T1]{fontenc}
\usepackage{graphicx}
\usepackage{subcaption}

\usepackage[shortcuts,acronym]{glossaries}
\glsdisablehyper
\newacronym{vm}{VM}{Virtual Machine}
\newacronym{gui}{GUI}{Graphical User Interface}
\newacronym{ebpf}{eBPF}{extended Berkeley Packet Filter}
\newacronym{cli}{CLI}{Command Line Interface}
\newacronym{csv}{CSV}{Comma-Separated Values}
\newacronym{des}{DES}{Discrete-Event Simulation}
\newacronym{lkm}{LKM}{Loadable Kernel Module}
\newacronym{gso}{GSO}{Generic Segmentation Offload}
\newacronym{edfq}{EDFQ}{Earliest Deadline First Queue}
\newacronym{fpu}{FPU}{Floating Point Unit}
\newacronym{pppoe}{PPPoE}{Point-to-Point Protocol over Ethernet}
\newacronym{qdisc}{qdisc}{queuing discipline}
\newacronym{aqm}{AQM}{Active Queue Management}

\usepackage{listings}
\usepackage[dvipsnames]{xcolor}
\definecolor{background}{HTML}{F5F5F5}
\definecolor{delim}{RGB}{20,105,176}
\colorlet{numb}{magenta!60!black}
\colorlet{punct}{red!60!black}

\lstdefinelanguage{cli}{
    basicstyle=\footnotesize\ttfamily,
    numbers=left,
    numberstyle=\scriptsize,
    stepnumber=1,
    numbersep=8pt,
    showstringspaces=false,
    breaklines=true,
    frame=lines,
    xleftmargin=5.0ex,
    backgroundcolor=\color{background},
    identifierstyle=\color{black},
    stringstyle=\color{black},
    captionpos=b,
    abovecaptionskip=0.6em
}

\copyrightyear{2025}
\acmYear{2025}

\usepackage{float}
\floatstyle{plain}
\newfloat{listing}{t}{}
\floatname{listing}{Listing}

\begin{document}

\title{\textit{TheaterQ}: A Qdisc for Dynamic Network Emulation}

\author{Martin Ottens \href{https://orcid.org/0009-0003-4257-0087}{\includegraphics[scale=0.04]{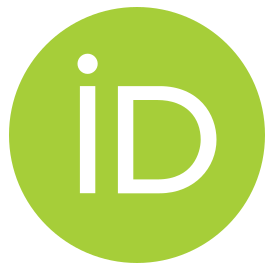}}{}}
\affiliation{%
  \institution{Friedrich-Alexander-Universität}
  \city{Erlangen-Nürnberg}
  \country{Germany}}
\email{martin.ottens@fau.de}

\author{Kai-Steffen Hielscher \href{https://orcid.org/0000-0002-2051-0660}{\includegraphics[scale=0.04]{figures/orcid.png}}{}}
\affiliation{%
  \institution{Friedrich-Alexander-Universität}
  \city{Erlangen-Nürnberg}
  \country{Germany}}
\email{kai-steffen.hielscher@fau.de}

\author{Reinhard German \href{https://orcid.org/0000-0002-9071-4802}{\includegraphics[scale=0.04]{figures/orcid.png}}{}}
\affiliation{%
  \institution{Friedrich-Alexander-Universität}
  \city{Erlangen-Nürnberg}
  \country{Germany}}
\email{reinhard.german@fau.de}

\renewcommand{\shortauthors}{Ottens et al.}

\begin{abstract}
    TheaterQ is a Linux \ac{qdisc} designed for dynamic network emulation, addressing the limitations of static parameters in traditional tools like NetEm. 
    By utilizing Trace Files containing timelines with network characteristics, TheaterQ achieves high-accuracy emulation of dynamic networks without involving the userspace and allows for resolutions of characteristic updates of up to 1~µs. 
    Features include synchronization across mutliple \ac{qdisc} instances and handling of delays, bandwidth, packet loss, duplication, and reordering. 
    Evaluations show TheaterQ's accuracy and its comparable performance to existing tools, offering a flexible solution for modern communication protocol development.
    TheaterQ is available as open-source software under the GPLv2 license.\footnote{\url{https://github.com/cs7org/TheaterQ}\label{fn:oss_link}}
\end{abstract}


\maketitle

\section{INTRODUCTION}
\label{sec:introduction}
Network emulation is an essential tool in protocol development, testing, and performance evaluation.
Testbeds currently used in research often only utilize static parameters for network emulation, e.g., a delay and bandwidth limit are selected once and not changed during an experiment.
A widely used tool for this purpose is \textit{NetEm}~\cite{hemminger2005}, which is included in the Linux kernel of many distributions.
However, when faced with emerging communication technologies, such as satellite mega constellations, networks have very dynamic characteristics, especially during longer observation periods. 
An example of this is the sudden delay changes that can occur during a satellite handover.

For this reason, different works have proposed approaches to emulate networks with dynamic characteristics.
Tian~et~al. are using \acs{ebpf} programs to emulate loss and delays on a packet-accurate level~\cite{tian2024}.
They demonstrated that their approach can reach high precision; however, since their traces contain delay and losses for each individual packet, it cannot be used with arbitrary traffic.
\textit{Link 'em} by Schütz~et~al. uses a modified version of NetEm to replay such packet-accurate loss traces~\cite{schuetz2019}.
Their approach also allows for the integration of more sophisticated loss models, making it suitable for use with arbitrary traffic.
Link 'em also only considers packet loss and delay; there is no functionality for dynamic changes in queue capacity or bandwidth, for example.

Other approaches provide link emulation functionality from within the userspace.
\textit{FlowEmu} by Stolpmann~et~al. provides a simple \ac{gui} to easily configure queues, bandwidths, losses, and delays.
The traffic is routed through the application that performs the emulation via raw sockets~\cite{stolpmann2021}.
Comparable to the other approaches, FlowEmu can only be used for emulating links with static characteristics or to replay packet-accurate traces.
Recently, Ohs~et~al. published \textit{PhantomLink}, a userspace application that works similarly to FlowEmu, but is configured using \ac{csv} files instead of a \ac{gui}~\cite{ohs2025}.
A significant difference is that these \ac{csv} files contain a timeline of link characteristics, where different characteristics are applied to the link emulation at specific times.
This allows for emulating links or network paths with dynamic characteristics and arbitrary traffic.
The authors have tailored PhantomLink especially for use in the area of satellite constellation emulation.

\begin{figure}[tb!]
    \centerline{\includegraphics[width=0.99\linewidth]{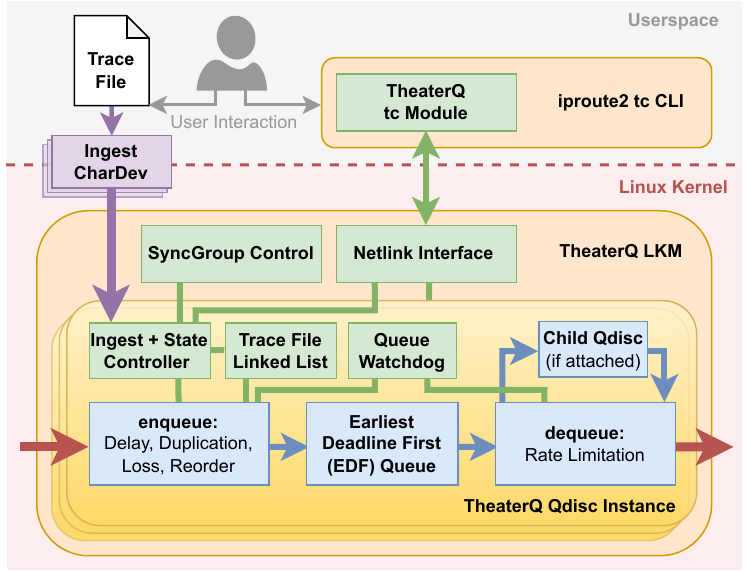}}
    \caption{Architecture of the \textit{TheaterQ} \acs{lkm}.}
    \label{fig:architecture}
\end{figure}

For an evaluation of their approach, the authors of PhantomLink have used a simplified simulation to obtain the parameters stored in the \ac{csv} file.
In previous work, we have described an approach for leveraging \acp{des} to export the dynamic characteristics of an end-to-end path to comparable \ac{csv} files, called \textit{Trace Files}~\cite{ottens2025-2}.
An example of such a Trace File can be found in Listing~\ref{lst:trace_file_example}.
In this previous work, we used a userspace wrapper that controls NetEm via its \textit{iproute2} \ac{cli} to apply the changing path characteristics to link emulation.
However, we encountered some challenges: With frequent updates of the characteristics, e.g., in millisecond intervals, a userspace application could delay updates, limiting reproducibility.
Also, for a decent link emulation, the forward and return links of a network are expected to have slightly different characteristics.
We therefore decided to use a Trace File for each path; however, our previous approach was unable to perfectly synchronize the playback of both.
We have also noticed limitations of NetEm itself: For example, the implementation does not allow implicit packet reordering (e.g., due to significant changes in delay) when a bandwidth limitation is configured.

To solve these challenges, we developed \textit{TheaterQ}, a Linux \acl{qdisc} for dynamic link emulation.
TheaterQ conducts link emulation as if directing a play, applying scripted changes provided in the Trace Files just in time.
Once configured, it should be able to replay Trace Files without additional support from the userspace, limiting the impact that other processes on the emulation system can have on the accuracy.
TheaterQ should also provide functionality to synchronize replays of Trace Files across different \ac{qdisc} instances (e.g., on different interfaces) and the setup should be easy and versatile, allowing integration into various emulation testbed setups.

This paper is structured as follows: Section~\ref{sec:implementation} describes the architecture of TheaterQ and presents an overview of its features. 
In Section~\ref{sec:usage}, a short usage guide is provided based on a simple example, which we also used for an evaluation.
A conclusion is provided in Section~\ref{sec:conclusion}.

\section{IMPLEMENTATION \& FEATURES}
\label{sec:implementation}
The TheaterQ \ac{qdisc} is provided by an \ac{lkm} and is based on the general implementation of NetEm.
An overview of the module's architecture is presented in Figure~\ref{fig:architecture}.
The \ac{lkm} can provide multiple \ac{qdisc} instances, each instance is located in the egress path of a network interface.
The instantiation and general configuration of a TheaterQ \ac{qdisc} is done from within the userspace via an \textit{iproute2} tc module.

\subsection{Operational Stages}
\begin{figure}[tb!]
    \centerline{\includegraphics[width=0.99\linewidth]{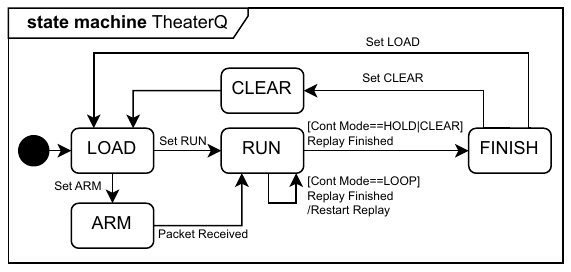}}
    \caption{Simplified state machine for a TheaterQ instance.}
    \label{fig:state_machine}
\end{figure}

Each TheaterQ \ac{qdisc} instance can be in one of five different stages; a simplified state diagram is shown in Figure~\ref{fig:state_machine}.
Initially, it starts in the \texttt{LOAD} stage, during which it behaves transparently (i.e., it does not emulate any characteristics), and Trace Files can be ingested via a character device opened for each instance.
After loading a Trace File, the replay can be started directly by setting the instance to the \texttt{RUN} stage.
Alternatively, setting the instance to the \texttt{ARM} stage will allow the instance to start the replay when the first packet is processed.
When the replay is finished, e.g., all entries in the Trace File were processed, the \textit{Continue Mode} (cont) will define how TheaterQ proceeds:
\begin{itemize}
    \item In the \texttt{HOLD} mode, the last Trace File values will be kept applied, and the instance transitions to the \texttt{FINISH} stage.
    \item In the \texttt{CLEAN} mode, the instance will behave transparently again, also transitioning to the \texttt{FINISH} stage.
    \item With the \texttt{LOOP} mode, the replay will restart at the beginning of the Trace File.
\end{itemize}
The user can clear all stored Trace File values by setting the instance to the \texttt{CLEAR} stage.
Setting the instance manually to the \texttt{LOAD} stage will stop the replay.
In this case, additional Trace File values can always be appended.

\subsection{Trace File Replay}
Each TheaterQ instance opens a character device used to ingest Trace Files in a specific \ac{csv} format.
Character devices are used instead of Netlink sockets or similar approaches due to their ease of use: Any program and even bash redirections can be used to ingest Trace Files.
A TheaterQ instance can be configured to accept an \texttt{EXTENDED} format as well as a \texttt{SIMPLE} format, where some parameters are omitted that are often not required.
Trace File entries are stored in a linked list within the \ac{qdisc} instance.

When an instance transitions into the \texttt{RUN} stage, the first entry of the Trace File is activated.
Each Trace File entry has a \texttt{keep\_us} (cf. Listing~\ref{lst:trace_file_example}) value, that defines the time after which it switches to the subsequent entry.
On arrival of a new packet, TheaterQ checks if current entry has expired, and performs a list walk to find the next Trace File value that should be active during this time.
This allows to have the correct characteristics applied for each packet by its time of arrival, without any time drifts.
With high update intervals and long periods between packets, list walks could take some time, possibly delaying packets.
With realistic scenarios, the runtime is negligible: With 100 link characteristic updates per second and 100 packets per seconds arriving, in average each packet will cause a list walk that sets the active entry pointer one list entry forward.
When the \texttt{LOOP} continue mode is selected, complete list walks are skipped whenever possible; this optimization is beneficial with larger intervals between successive packets.
Once the Trace File replay has started, the userspace is no longer involved, as everything is handled inside the kernel module.

\begin{listing*}[!t]
\begin{lstlisting}[language=cli,firstnumber=1,caption=Required commands to configure and start a path emulation with two TheaterQ instances in a SyncGroup.,label=lst:cli_example]
 tc qdisc add dev <@\textcolor{BrickRed}{eth1}@> root handle <@\textcolor{BrickRed}{1}@> theaterq ingest <@\textcolor{OliveGreen}{SIMPLE}@> cont <@\textcolor{OliveGreen}{LOOP}@> syncgroup <@\textcolor{OliveGreen}{1}@> <@\textcolor{gray}{\# Stage initially is LOAD}@>
 tc qdisc add dev <@\textcolor{BrickRed}{eth2}@> root handle <@\textcolor{BrickRed}{1}@> theaterq ingest <@\textcolor{OliveGreen}{SIMPLE}@> cont <@\textcolor{OliveGreen}{LOOP}@> syncgroup <@\textcolor{OliveGreen}{1}@>
 cat forward_trace.csv > <@\textcolor{BrickRed}{/dev/theaterq\textbackslash{}:eth1\textbackslash{}:1\textbackslash{}:0}@> <@\textcolor{gray}{\# Ingest Trace Files to qdisc instances}@>
 cat return_trace.csv > <@\textcolor{BrickRed}{/dev/theaterq\textbackslash{}:eth2\textbackslash{}:1\textbackslash{}:0}@>
 tc qdisc change dev <@\textcolor{BrickRed}{eth1}@> handle <@\textcolor{BrickRed}{1}@> theaterq stage <@\textcolor{OliveGreen}{ARM}@> <@\textcolor{gray}{\# First packet leaving via eth1 will start both syncgroup members}@>
\end{lstlisting}
\end{listing*}

\subsection{SyncGroups}
To synchronously start the Trace File replay within different TheaterQ instances for multiple network interfaces on the same system, a \textit{SyncGroup} feature is provided.
Each \ac{qdisc} instance can be a member of one SyncGroup.
When one member of the SyncGroup is set to the \texttt{RUN} stage, and the replay is started, all other members will also transition to the \texttt{RUN} stage.
For this to work, the other members must be ready, e.g., a Trace File must be ingested.
This, of course, also applies to the \texttt{ARM} stage: The first packet received by any member instance of a SyncGroup that is in the \texttt{ARM} stage will set all members to the \texttt{RUN} stage.
Only the start of a replay is synchronized: After this, all members can have different update times in their Trace Files, different Trace File lengths, and different continue modes can be chosen.

\subsection{Characteristics}
Depending on the selected Trace File format, TheaterQ can emulate different characteristics.
NetEm provides complex functionalities for modelling different aspects, like loss or delay with correlations and statistical models, or even for detailed emulation of slotted networks.
Since TheaterQ can provide frequent updates of the link characteristics, we avoided this complexity, concentrating on simpler parameters.
By using zero values in the Trace Files for any given characteristic, TheaterQ will handle this characteristic as transparently as possible.
All characteristics are applied when a packet is received by the TheaterQ instance (during \textit{enqueue} in Figure~\ref{fig:architecture}), thus a change of the active Trace File entry while a packet is in the queue has no effect on that packet anymore.
The different characteristics are presented in the following.

\subsubsection{Rate Limit}
A rate limit in bits per second can be selected to emulate a bottleneck.
To account for additional packet encapsulations that increase its size, e.g., \acs{pppoe}, a static packet overhead can be configured.
When enabled, TheaterQ can automatically segment \ac{gso} packets~\cite{xu2006} into real Ethernet packets, which increases the accuracy and is also helpful for the accuracy of the other characteristics.

\subsubsection{Packet Delay}
Packets can be delayed by a specific time.
Additionally, a standard deviation can describe the jitter of this delay.
We assume that such fine-grained jitters are not required just for packet delays, since frequent Trace File entry updates should be sufficient.
However, as well as sudden changes in the delays, high jitters can lead to implicit packet reordering.
Since NetEm cannot emulate this behavior correctly when a rate limit and delay are configured, we have modified its queue implementation.
Our \ac{edfq} tracks the earliest time a packet can be sent alongside its transmission duration, accounting for the delay and rate limit.
The dequeue function of the \ac{qdisc} keeps track of until which time the link is considered busy, e.g., when the next packet can be transmitted.
When the link is available again, a watchdog will trigger the \textit{dequeue} function, transmitting the packet from the queue with the earliest deadline while considering the stored transmission time.
The \ac{edfq} allows for arbitrary packet reordering, while leveraging the performance of the original red-black tree of NetEm's \textit{tfifo}.

\subsubsection{Queue Limits}
Each TheaterQ instance keeps track of the size of the \ac{edfq} in bytes and packets.
An upper limit for the queue can be defined, and during instance setup, the user can select if a byte or packet limit should be used.
Whenever the queue has insufficient capacity at the arrival time of a packet, it is dropped.

\subsubsection{Packet Loss}
A packet loss ratio is used pseudo-randomly drop packets before they are enqueued to the \ac{edfq}.
In the Trace File, a scaled 32-bit value is used as a drop probability (0 = 0\%, \texttt{UINT32\_MAX} = 100\%), since the kernel does not have access to the \ac{fpu}.
A drop probability of 100\% can be used to emulate a temporary connection loss.

\subsubsection{Duplication}
TheaterQ can perform pseudo-random duplications of packets.
For this, two values are used: The duplication probability, a scaled 32-bit value, and a delay offset.
The delay offset describes how long the duplicate is additionally delayed before transmission.
This can be useful to reorder duplicated packets inside a flow.
A duplicated packet is enqueued like a regular packet; thus, delays and limits apply.

\subsubsection{Route Identifiers}
In some cases, implicit packet reordering due to delay changes or a high jitter are unwanted, e.g., packets transmitted via the same network route will generally not encounter reordering.
To control this behavior, a route identifier can be specified for each Trace File entry, comparable to the approach presented by Ohs~et~al.~\cite{ohs2025}.
Packets transmitted while the same route identifier is active are strictly ordered, possibly overwriting calculated delays.
TheaterQ internally keeps track of the timestamp the last packets is scheduled for transmission for each route identifier.
The route identifier 0 is reserved and can be used to enable implicit packet reordering while the Trace File entry is active.

\section{USAGE \& EXAMPLE}
\label{sec:usage}

\begin{figure}[tb!]
    \centerline{\includegraphics[width=0.99\linewidth]{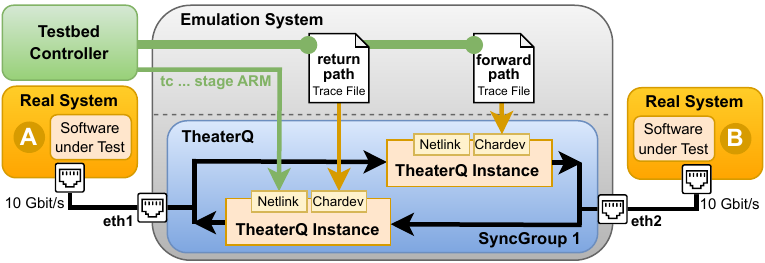}}
    \caption{Example setup using an emulation system with two network interfaces between test systems.}
    \label{fig:setup}
\end{figure}

\begin{figure*}[t!]
  \centering
  \begin{subfigure}{.49\textwidth}
    \centering
    \includegraphics[width=.99\linewidth]{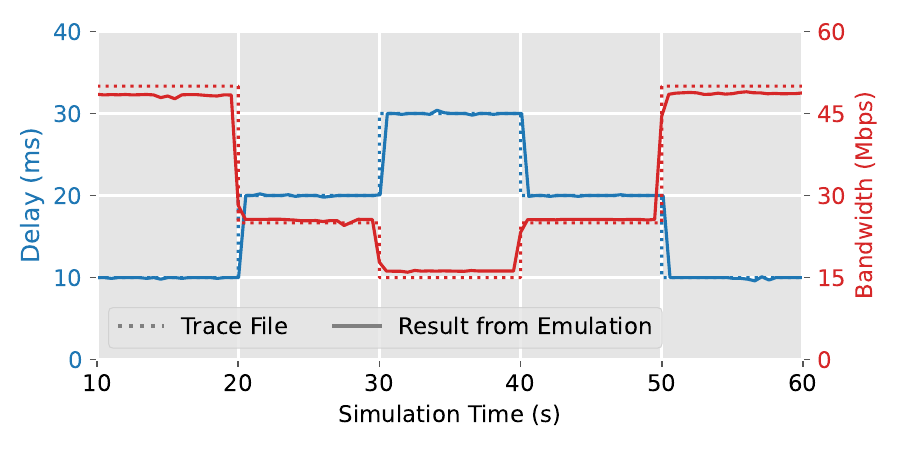}
    \caption{Simple scenario, comparable with the evaluation scenario\\presented in~\cite{ottens2025-2}.
             The related Trace File is shown in Listing~\ref{lst:trace_file_example}.}
    \label{fig:results-simple}
  \end{subfigure}%
  \begin{subfigure}{.49\textwidth}
    \centering
    \includegraphics[width=.99\linewidth]{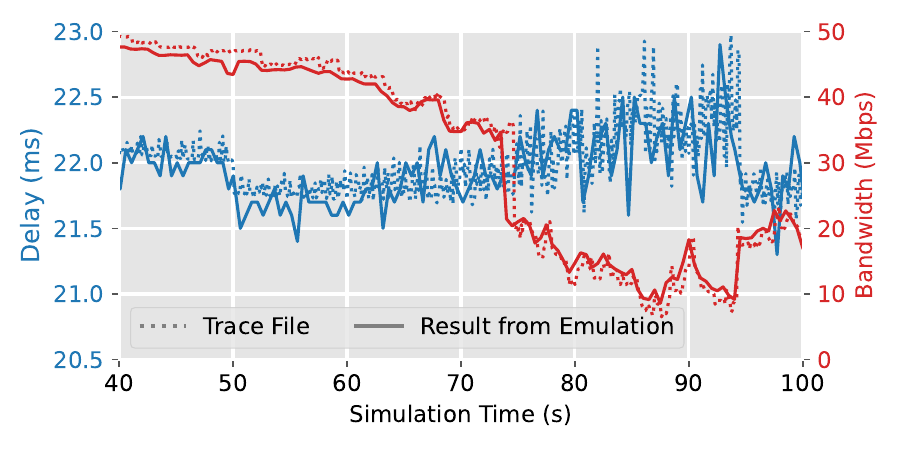}
    \caption{Complex scenario with a Trace File of an end-to-end path exported from a satellite simulator with 10~ms characteristic updates.}
    \label{fig:results-complex}
  \end{subfigure}%
  \caption{Comparison of the values from the Trace Files and actual Layer 2-throughput using UDP traffic and ICMP ping RTT measurements in the emulation environment. Only bandwidth limitations and delay characteristics were applied on the forward path, e.g., only one TheaterQ instance was used.}
  \label{fig:results}
\end{figure*}

Before TheaterQ can be used, the Trace Files describing the link characteristics must be generated.
Trace Files could be obtained in several ways: With a high-level simulation~\cite{ohs2025}, using a modified \ac{des}~\cite{ottens2025-2}, with measurements from real networks, or they could be hand-crafted for specific scenarios.
Subsequently, an emulation environment must be chosen.
Due to TheaterQ's flexibility, various setups are possible.
During development, we have used a setup based on \acp{vm}~\cite{ottens2025-3}; however, TheaterQ can be integrated into namespace-based setups, such as MiniNet~\cite{mininet2022}, or used with \textit{Docker} containers.
It is also possible to use TheaterQ in physical testbeds, comparable to the setup used by Link 'em~\cite{schuetz2019}.
Like NetEm and other packet schedulers, TheaterQ can only be used on the egress path of a network interface.
Workarounds for ingress traffic are possible, e.g., using ifb interfaces\footnote{Intermediate Functional Block, see \url{https://wiki.linuxfoundation.org/networking/ifb}}.
Comparable to NetEm, TheaterQ also supports a child \ac{qdisc}.
Packets dequeued from the \ac{edfq}, thus after any delays and limitations were processed, could be enqueued to this child \ac{qdisc}.
Since packets could be dropped due to full queues before the child has processed them, usability of \ac{aqm} \acp{qdisc} as a child are limited.

\subsection{Example Setup}
In the following, we are describing the setup shown in Figure~\ref{fig:setup}, which is based on \acp{vm}.
Two test systems, $A$ and $B$, are connected via an emulation system, which has two network interfaces.
The emulation system performs the link emulation transparently for the test systems, as both interfaces are bridged.
Traffic from $A$ to $B$ is leaving the emulation system via the second interface (cf. \textit{eth2} in Figure~\ref{fig:setup}): At this point, a TheaterQ instance is installed, performing the link emulation for the forward path.
A second instance is used on the other interface for the return path (cf. \textit{eth1} in Figure~\ref{fig:setup}).
In this case, the replay of both Trace Files must be synchronized to achieve the desired results.

\begin{listing}[b!]
\begin{lstlisting}[language=cli,firstnumber=1,caption=Example of a Trace File in the \texttt{SIMPLE} format with updates every 10~seconds. See Figure~\ref{fig:results-simple} for a visualization.,label=lst:trace_file_example]
<@\textcolor{BrickRed}{keep\_us,\phantom{0}}@> <@\textcolor{OliveGreen}{delay\_us,}@> <@\textcolor{BlueViolet}{rate\_bps,}@> <@\textcolor{Plum}{loss\_prob,}@> <@\textcolor{Bittersweet}{q\_limit}@>
<@\textcolor{BrickRed}{10000000,}@> <@\textcolor{OliveGreen}{10000,\phantom{000}}@> <@\textcolor{BlueViolet}{50000000,}@> <@\textcolor{Plum}{0,\phantom{00000000}}@> <@\textcolor{Bittersweet}{300}@>
<@\textcolor{BrickRed}{10000000,}@> <@\textcolor{OliveGreen}{20000,\phantom{000}}@> <@\textcolor{BlueViolet}{25000000,}@> <@\textcolor{Plum}{0,\phantom{00000000}}@> <@\textcolor{Bittersweet}{300}@>
<@\textcolor{BrickRed}{10000000,}@> <@\textcolor{OliveGreen}{30000,\phantom{000}}@> <@\textcolor{BlueViolet}{15000000,}@> <@\textcolor{Plum}{0,\phantom{00000000}}@> <@\textcolor{Bittersweet}{300}@>
<@\textcolor{BrickRed}{10000000,}@> <@\textcolor{OliveGreen}{20000,\phantom{000}}@> <@\textcolor{BlueViolet}{25000000,}@> <@\textcolor{Plum}{0,\phantom{00000000}}@> <@\textcolor{Bittersweet}{300}@>
<@\textcolor{BrickRed}{10000000,}@> <@\textcolor{OliveGreen}{10000,\phantom{000}}@> <@\textcolor{BlueViolet}{50000000,}@> <@\textcolor{Plum}{0,\phantom{00000000}}@> <@\textcolor{Bittersweet}{300}@>
\end{lstlisting}
\end{listing}

TheaterQ is installed by loading the \ac{lkm} and setting up the iproute2 tc extension.
To configure TheaterQ for the previously described setup, only the commands shown in Listing~\ref{lst:cli_example} are required.
While adding the \ac{qdisc} instances to the interfaces, initial settings, like SyncGroup membership and continue mode, are selected.
The TheaterQ instances are now in the \texttt{LOAD} stage, and the Trace Files can be written to the character device.
This character device is named after the interface, the instance is attached to, and the instance's major handle.

To begin with an experiment, one of the instances is set to the \texttt{ARM} stage.
In this case, the first packet processed by this TheaterQ instance will initiate the Trace File replay for both instances in the SyncGroup, setting them to the \texttt{RUN} stage.
During an experiment, the show and statistics commands from iproute2 tc can be used to monitor the playback process and the characteristics currently active.

\subsection{Evaluation}
To demonstrate TheaterQ's accuracy, we have performed a small evaluation using two different Trace Files.
The emulation setup is comparable to that shown in Figure~\ref{fig:setup}, but only one TheaterQ instance for the forward path is used.
We measured the Layer 2 throughput achieved by a UDP flood as well as the ICMP ping RTT, while no additional traffic was present in the emulation environment.

Listing~\ref{lst:trace_file_example} shows the first Trace File.
This Trace File was hand-crafted, but is oriented on the evaluation topology presented in our previous work~\cite{ottens2025-2}.
Figure~\ref{fig:results-simple} compares the results from the measurements with the values from the Trace File.
Apart from the minor expected deviations, both measurements achieve exactly the expected results.

In the case of Figure~\ref{fig:results-complex}, a Trace File from a complex satellite simulation was used.
This simulation yields very dynamic path characteristics with updated values in the Trace Files every 10~ms.
Even with such frequent changes, TheaterQ can reproduce the characteristics contained in the Trace Files.

We have compared the steady-state performance of TheaterQ, e.g., with changes in the characteristics, with NetEm: Both \acp{qdisc} perform identically.
The accuracy of NetEm is limited by the kernel's scheduler, timer resolutions, and system load, both of this is also the case for TheaterQ.

\section{CONCLUSION}
\label{sec:conclusion}
In this paper, we presented \textit{TheaterQ}, a Linux queuing discipline for emulating links with dynamic characteristics, as required to keep up with novel communication technologies during protocol development.
TheaterQ is partially based on NetEm, thus easy to use, and flexible integrable in different emulation testbeds.
A timeline of varying link characteristics, referred to as Trace Files, is ingested into TheaterQ instances via a character device in a \ac{csv} format, eliminating the need for additional programs.
After initial configuration and Trace File ingestion, TheaterQ handles the replay without involving the userspace.
The SyncGroup feature allows synchronous starting of Trace File replays in different TheaterQ instances on the same system, which is helpful for complex emulation setups that involve multiple network interfaces.

TheaterQ is currently able to emulate delays with jitter, bandwidth limits, packet drops and duplications, queue size limits, as well as packet reordering.
In future work, additional statistical models could be integrated, for example, to describe correlated packet loss with even higher precision.

In an evaluation, we demonstrated the performance of TheaterQ to reproduce the link characteristics from a Trace File accurately.
It performs as well as NetEm does when used with a comparable set of features, even if TheaterQ has some significant implementation differences, e.g., to allow correct packet reordering.
TheaterQ is available under a GPLv2 license as open-source software\footref{fn:oss_link}, and we believe it can become a valuable tool in future protocol development and performance evaluation.

\bibliographystyle{ACM-Reference-Format}
\bibliography{bibliography}


\begin{thebibliography}{9}


\ifx \showCODEN    \undefined \def \showCODEN     #1{\unskip}     \fi
\ifx \showISBNx    \undefined \def \showISBNx     #1{\unskip}     \fi
\ifx \showISBNxiii \undefined \def \showISBNxiii  #1{\unskip}     \fi
\ifx \showISSN     \undefined \def \showISSN      #1{\unskip}     \fi
\ifx \showLCCN     \undefined \def \showLCCN      #1{\unskip}     \fi
\ifx \shownote     \undefined \def \shownote      #1{#1}          \fi
\ifx \showarticletitle \undefined \def \showarticletitle #1{#1}   \fi
\ifx \showURL      \undefined \def \showURL       {\relax}        \fi
\providecommand\bibfield[2]{#2}
\providecommand\bibinfo[2]{#2}
\providecommand\natexlab[1]{#1}
\providecommand\showeprint[2][]{arXiv:#2}

\bibitem[Contributors(2022)]%
        {mininet2022}
\bibfield{author}{\bibinfo{person}{Mininet~Project Contributors}.} \bibinfo{year}{2022}\natexlab{}.
\newblock \bibinfo{title}{{Mininet} -- {An} {Instant} {Virtual} {Network} on your {Laptop} (or other {PC})}.
\newblock
\urldef\tempurl%
\url{https://mininet.org/}
\showURL{%
\tempurl}


\bibitem[Hemminger et~al\mbox{.}(2005)]%
        {hemminger2005}
\bibfield{author}{\bibinfo{person}{Stephen Hemminger} {et~al\mbox{.}}} \bibinfo{year}{2005}\natexlab{}.
\newblock \showarticletitle{{Network} emulation with {NetEm}}. In \bibinfo{booktitle}{\emph{Linux Conf AU}}, Vol.~\bibinfo{volume}{5}. \bibinfo{address}{Canberra, Australia}, \bibinfo{pages}{2005}.
\newblock


\bibitem[Ohs et~al\mbox{.}(2025)]%
        {ohs2025}
\bibfield{author}{\bibinfo{person}{Robin Ohs}, \bibinfo{person}{Gregory~F. Stock}, \bibinfo{person}{Juan~A. Fraire}, \bibinfo{person}{Holger Hermanns}, {and} \bibinfo{person}{Andreas Schmidt}.} \bibinfo{year}{2025}\natexlab{}.
\newblock \showarticletitle{{PhantomLink}: {Emulating} {Virtual} {End-to-End} {Links} on {Ground} and in {Orbit}}. In \bibinfo{booktitle}{\emph{Proceedings of the 2025 Applied Networking Research Workshop}} (Madrid, Spain) \emph{(\bibinfo{series}{ANRW '25})}. \bibinfo{publisher}{Association for Computing Machinery}, \bibinfo{address}{New York, NY, USA}, \bibinfo{pages}{39–46}.
\newblock
\showISBNx{9798400720093}
\href{https://doi.org/10.1145/3744200.3744758}{doi:\nolinkurl{10.1145/3744200.3744758}}


\bibitem[Ottens et~al\mbox{.}(2025a)]%
        {ottens2025-3}
\bibfield{author}{\bibinfo{person}{Martin Ottens}, \bibinfo{person}{Jörg Deutschmann}, \bibinfo{person}{Kai-Steffen Hielscher}, {and} \bibinfo{person}{Reinhard German}.} \bibinfo{year}{2025}\natexlab{a}.
\newblock \showarticletitle{{Proto2Testbed}: {Towards} an {Integrated} {Testbed} for {Evaluating} {End-to-End} {Security} {Protocols} in {Satellite} {Constellations}}. In \bibinfo{booktitle}{\emph{2025 12th Advanced Satellite Multimedia Systems Conference and the 18th Signal Processing for Space Communications Workshop (ASMS/SPSC)}}. \bibinfo{pages}{1--8}.
\newblock
\href{https://doi.org/10.1109/ASMS/SPSC64465.2025.10946051}{doi:\nolinkurl{10.1109/ASMS/SPSC64465.2025.10946051}}


\bibitem[Ottens et~al\mbox{.}(2025b)]%
        {ottens2025-2}
\bibfield{author}{\bibinfo{person}{Martin Ottens}, \bibinfo{person}{Kai-Steffen Hielscher}, {and} \bibinfo{person}{Reinhard German}.} \bibinfo{year}{2025}\natexlab{b}.
\newblock \showarticletitle{From {Simulation} to {Emulation}: {A} {System} {Design} for {Real-Time} {Replay} of {Simulated} {Network} {Path} {Characteristics}}. In \bibinfo{booktitle}{\emph{Proceedings of the 2025 International Conference on Ns-3}} (Osaka, JP) \emph{(\bibinfo{series}{ICNS3 '25})}. \bibinfo{publisher}{Association for Computing Machinery}, \bibinfo{address}{New York, NY, USA}, \bibinfo{pages}{62–69}.
\newblock
\showISBNx{9798400715174}
\href{https://doi.org/10.1145/3747204.3747206}{doi:\nolinkurl{10.1145/3747204.3747206}}


\bibitem[Schütz et~al\mbox{.}(2019)]%
        {schuetz2019}
\bibfield{author}{\bibinfo{person}{Bertram Schütz}, \bibinfo{person}{Stefanie Thieme}, \bibinfo{person}{Nils Aschenbruck}, \bibinfo{person}{Leonhard Brüggemann}, \bibinfo{person}{Alexander Ditt}, \bibinfo{person}{Dominic Laniewski}, {and} \bibinfo{person}{Dennis Rieke}.} \bibinfo{year}{2019}\natexlab{}.
\newblock \showarticletitle{{Link 'em}: {An} {Open} {Source} {Link} {Emulation} {Bridge} for {Reproducible} {Networking} {Research}}. In \bibinfo{booktitle}{\emph{2019 International Conference on Networked Systems (NetSys)}}. \bibinfo{pages}{1--3}.
\newblock
\href{https://doi.org/10.1109/NetSys.2019.8854509}{doi:\nolinkurl{10.1109/NetSys.2019.8854509}}


\bibitem[Stolpmann and Timm-Giel(2021)]%
        {stolpmann2021}
\bibfield{author}{\bibinfo{person}{Daniel Stolpmann} {and} \bibinfo{person}{Andreas Timm-Giel}.} \bibinfo{year}{2021}\natexlab{}.
\newblock \showarticletitle{{FlowEmu}: {An} open-source flow-based network emulator}.
\newblock \bibinfo{journal}{\emph{Electronic Communications of the EASST}}  \bibinfo{volume}{80} (\bibinfo{year}{2021}).
\newblock
\href{https://doi.org/10.14279/tuj.eceasst.80.1141}{doi:\nolinkurl{10.14279/tuj.eceasst.80.1141}}


\bibitem[Tian et~al\mbox{.}(2024)]%
        {tian2024}
\bibfield{author}{\bibinfo{person}{Weibiao Tian}, \bibinfo{person}{Ye Li}, \bibinfo{person}{Jinwei Zhao}, \bibinfo{person}{Sheng Wu}, {and} \bibinfo{person}{Jianping Pan}.} \bibinfo{year}{2024}\natexlab{}.
\newblock \showarticletitle{An {eBPF-Based} {Trace-Driven} {Emulation} {Method} for {Satellite} {Networks}}.
\newblock \bibinfo{journal}{\emph{IEEE Networking Letters}} \bibinfo{volume}{6}, \bibinfo{number}{3} (\bibinfo{year}{2024}), \bibinfo{pages}{188--192}.
\newblock
\href{https://doi.org/10.1109/LNET.2024.3472034}{doi:\nolinkurl{10.1109/LNET.2024.3472034}}


\bibitem[Xu(2006)]%
        {xu2006}
\bibfield{author}{\bibinfo{person}{Herbert Xu}.} \bibinfo{year}{2006}\natexlab{}.
\newblock \bibinfo{title}{{GSO}: {Generic} {Segmentation} {Offload}}.
\newblock \bibinfo{howpublished}{\url{https://lwn.net/Articles/188489/}}.
\newblock


\end{thebibliography}

\small{All links were last accessed on \today{}.}

\end{document}